\documentclass[aps,prl,nofootinbib,twocolumn,letterpaper]{revtex4}

\begin{document}

\title{Reply to  ``Comment on 
Testing Planck-Scale Gravity with Accelerators''}

\author{Vahagn Gharibyan}
\email[]{vahagn.gharibyan@desy.de}

\affiliation{Deutsches Elektronen-Synchrotron DESY - D-22603 Hamburg}

\maketitle

In 2012 Letter Ref.\cite{Gharibyan:2012gp}  I have 
discussed a possibility to test Planck-scale space 
birefringence and refractivity for a leading order 
energy dependent photon dispersion model.

In a recent Comment~\cite{Kalaydzhyan:2016jqy} 
Kalaydzhyan questions correctness of 
Ref.\cite{Gharibyan:2012gp} results, calling the method 
and conclusions wrong.
Arguments in the Comment, however, are based on misunderstanding
or incorrect assumptions making the claim invalid.
Below I will address all concerns raised in the Comment. 

1.
According to the Comment, the experimental results in the Letter,
obtained with HERA and SLC Compton beams, 
are excluded by previous non-observation of vacuum Cherenkov 
radiation at LEP. 
The statement, however, overlooks vacuum Cherenkov limits
quoted in the Letter. New calculation suggested in the Comment is 
inaccurate since it ignores energy dependence of the Planck-scale 
refractive index 
\begin{equation}
n=1+\zeta \frac{\omega}{M_P},
\label{nqg}
\end{equation}
with $\omega$, $M_P$ and $\zeta$ being the photon energy,
the Planck mass and the scaling constant respectively 
(Eq.(\ref{nqg}) is equivalent to the Letter's Eq.7   
for $\mathcal{O}((\omega/M_P)^2)$)
A correct approach is to substitute the $n$ in the vacuum Cherenkov 
formula (e.g. in Eq.(1) of the Comment or Ref.\cite{Gharibyan:2003fe}) 
by Eq.(\ref{nqg}) and obtain limits on $\zeta$. 
This has been done in the Letter where a
tighter-than-LEP astrophysical bound $\zeta<300$ is quoted
for 3~TeV electrons (see section "Current limits"). 
Its easy to see that the HERA and SLC refractivities
correspond to  
\hbox{$ \zeta = -1.6\cdot 10^{7} $} and  \hbox{$ \zeta = -2.2\cdot 10^{5} $}
values which are well below the limit exposed by non-observation of vacuum 
Cherenkov.
These numbers are directly obtained from the measured refractivities, 
photon energies, Eq.(1)
and the refractivity signs ($n< 1$) quoted in Ref.~\cite{Gharibyan:2012gp} 
and Ref.~\cite{Gharibyan:2003fe}.

2.
Next concern in the Comment is the electron's zero dispersion 
at the Planck-scale vacuum which is suggested to replace by something 
to reproduce a general relativistic term 
from the Comment's Eq.2, at classical limit $M_P \rightarrow  \infty$.
However, a closer look to this term
\begin{equation}
  n = 1+\frac{2GM}{c^2 R},
\label{ne}
\end{equation}
with the gravitational constant G, 
reveals a real gravitational field origin 
from a spherical mass $M$ and radius $R$.
Obviously, the real field refractivity in Eq.(\ref{ne}) can not enter 
to description of any vacuum whether classical or quantum-gravitational
since no vacuum can convert into real gravity (except at Big-Bang singularity). 
In case the refractivity in Eq.(\ref{ne}) is assigned to the Earth's field in 
the laboratory, the same term will have also the photon in the Eq.(\ref{nqg}) 
which eventually will drop from the final result according to the equivalence 
principle.   
Thus, the suggested general relativistic field analogy is not applicable for the 
vacuum which, in contrast to real gravitational field, differently couples
to photons and electrons because of charge and spin differences.
Additional concerns in the Comment about possible 
quantum gravity signals mimicked by equivalence principle
1\% violation could be handled by exploring energy
dependence at the Planck-scale vacuum. In case
the equivalence principle violation will persist 
to mimic also the energy dependence, the Planck vacuum 
could be separated from the real gravitational field 
by repeating the experiment at some (orbital) distance from 
the Earth. Thus, the Compton method has a potential to
access the quantum Planck-scale even in case of broken 
equivalence principle.

3.
The last critical remark in the Comment is a reference to competing
electromagnetic effects such as electron-beam and electron-vacuum chamber
interactions. 
In case the remark is addressing the 
HERA and SLC results, in Ref.\cite{Gharibyan:2003fe} all the essential 
electromagnetic backgrounds are quantified and included in the systematic 
error.
If, however,  the criticism is concerning future Planck scale vacuum 
Compton tests, a possible bad influence of the beam or the vacuum chamber
could be handled by experimental means - reducing the beam current or 
using non-conductive vacuum chamber. 

In conclusion, the criticisms in the Comment are theoretically and 
experimentally unjustified and I confirm the original results and
conclusions.

\end{document}